\documentclass[11pt,twoside,english]{article}
\usepackage[T1]{fontenc}
\usepackage[latin1]{inputenc}
\usepackage{graphicx}
\usepackage{amssymb}
\usepackage{babel}

\begin{document}

\title{Conjugation of cascades}

\author{Jesús San Martín${}^{a,b}$, Daniel Rodríguez-Pérez${}^{b}$}

\maketitle
$^{a}$ Departamento de Matemática Aplicada, E.U.I.T.I, Universidad
Politécnica de Madrid. Ronda de Valencia 3, 28012 Madrid Spain

$^{b}$ Departamento de Física Matemática y Fluidos, U.N.E.D. Senda
del Rey 9, 28040 Madrid Spain

Corresponding author: jsm@dfmf.uned.es

\begin{abstract}
Presented in this work are some results relative to sequences found
in the logistic equation bifurcation diagram, which is the unimodal
quadratic map prototype. All of the different saddle-node bifurcation
cascades, associated to every last appearance $p$-periodic orbit
($p=3,\,4,\,5,\dots$), can also be generated from the very Feigenbaum
cascade. In this way it is evidenced the relationship between both
cascades. The orbits of every saddle-node bifurcation cascade, mentioned
above, are located in different chaotic bands, and this determines
a sequence of orbits converging to every band-merging Misiurewicz
point. In turn, these accumulation points form a sequence whose accumulation
point is the Myrberg-Feigenbaum point. It is also proven that the
first appearance orbits in the $n$-chaotic band converge to the same
point as the last appearance orbits of the $(n+1)$-chaotic band.
The symbolic sequences of band-merging Misiurewicz points are computed
for any window.
\end{abstract}

\section{Introduction}

Nonlinear dynamical systems exhibit a rich variety of behaviors. Bifurcations,
chaos, patterns, phase transitions are among the most investigated
phenomena, and they have led to many discussions within scientific
literature. The former mentioned phenomena appears in a lot of dynamical
systems, and this is an indicative of the existence of underlaying
common structures whose origin claims to be uncovered.

To be able to undertake the study of a dynamical system, diverse tools
have been developed, which make the study of such problems easier,
by means of simplifications, classifications or through the identification
of universality classes among transitions from one behavior to another.

Near a hyperbolic fixed point, the well known Hartman-Gro{\ss}man
theorem states that the local phase portrait is topologically conjugate
to the phase portrait of the linearized system. However, when the
fixed point is a non-hyperbolic one, the center manifold must be considered.
The Center Manifold theorem allows one to generalize
the ideas of the Hartman-Gro{\ss}man theorem. It is deduced that
the flow, restricted to the center manifold, determines the topological
behavior close to a non-hyperbolic fixed point. In general, when the
flow is restricted to the center manifold the problem is simplified
quite a bit, yet the restricted flow itself can be very complicated.

The normal form theorem is the next mathematical tool that must be
used: a smooth vector field is substituted by a polynomial vector
field. Another technique very similar, in certain way, to the tool
stated above consists in constructing a Poincaré section. After that,
the Poincaré map is created, that is, a map from the Poincaré section
in itself. In dissipative systems this map contracts the Poincaré
section surface after each iteration, and the initial problem is thus
transformed into another problem with an attractor of lower dimensionality.

Fortunately, one-dimensional maps provide a good approximation to
many dynamical systems. Furthermore one-dimensional maps can be classified
into universality classes. That is why the bifurcation route to chaos,
discovered by Feigenbaum \cite{Feigenbaum78,Feigenbaum79} and characterized
by its universal scaling is so important. And for the same reason
the logistic equation\begin{equation}
x_{n+1}=f(x_{n})=\mu x_{n}(1-x_{n})\quad\mu\in[0,4],\; x_{n}\in[0,1]\label{eq:logistic}\end{equation}
is so intensively studied. The different behaviors of this map can
be extended to any quadratic one-dimensional map, because they are
topologically conjugated \cite{Milnor88}. Therefore, any dynamical
system ruled by a quadratic polynomial behaves as if it were ruled
by the logistic equation.

~

The goal of this paper is to show in the logistic equation the interesting
connection between the Feigenbaum cascade and the saddle-node cascade
\cite{SanMartin07}. Also this connection affects the global structures
of windows and Misiurewicz points \cite{Misiurewicz91} where chaotic
bands mix. Given the multitude of dynamical systems that exhibit Feigenbaum's
universality, it is very probable that they also show the new universal
behaviors described in this paper.

Periodic windows, apart from being an irreplaceable tool in system
dynamics, also play other roles and appear in different problems of
nonlinear dynamics. They can be used, by applying the Poincaré map,
to spot the windows of a dynamical systems which are ruled by ordinary
differential equations \cite{Churukian96}. Chaos control techniques
use periodic windows too and they use unstable periodic orbits inside
them \cite{Xu96,Bishop98}. Periodic windows are also useful to study
chaos-induced diffusion \cite{Blackburn98} which is affected by the
presence of proximate periodic windows. Therefore any contribution
aimed at the identification of the laws ruling periodic windows will
be useful in these related fields.

~

This paper is organized as follows:

In section \ref{sec:CofC}, it is proven that there is an analytic
relation among saddle-node bifurcation cascades \cite{SanMartin07}
and Feigenbaum cascade \cite{Feigenbaum78,Feigenbaum79}. The former
is derived from the latter. We will refer to this relation by \emph{conjugation
of cascades}. The section \ref{sec:CofC} is preceded by section \ref{sec:Definitions}
where later used definitions and theorems are introduced.

In the following section \ref{sec:1st-and-last}, the previous result
is used to prove that saddle-node bifurcation cascades associated
to last appearance orbits in the $(n+1)$-chaotic band and the ones
associated to first appearance orbits in the $n$-chaotic band converge
to the same Misiurewicz point where the $(n+1)$-chaotic and $n$-chaotic
bands mix.

Section \ref{sec:generalization} generalizes the former results,
which were only valid in the canonical window, to any periodic window.
In particular, we will also generalize the rule to get the saddle-node
bifurcation cascade in any window. As a spin-off, a geometrical interpretation
of the Derrida, Gervois and Pomeau (DGP) formula \cite{Derrida79}
is achieved.

In Section \ref{sec:disc-and-conclusions}, some potential applications
of the theorems to experimental research are indicated.

\section{\label{sec:Definitions}Definitions}

The following notation, definitions and theorems
will be referred to in what follows. Whenever they are based on or
copied from a previous work, the original reference is given.

\begin{description}
\item [{Definition~1.}] $S_{n}$ will denote a $n$-periodic symbolic
pattern. The symbols are $R$ or $L$, depending on wether the iterates
$x_{i}$ of a $n$-periodic orbit of the logistic map (\ref{eq:logistic})
are $x_{i}>\frac{1}{2}$ or $x_{i}<\frac{1}{2}$ ($i=1,2,\,3,\dots n$),
respectively \cite{MetropolisSteinStein73}. In particular, to determine
in an unique way the periodic sequence $S_{n}$, we will always take
as the first letter in the sequence that corresponding to the position
$x_{i}$ nearest to $1/2$, the rest of the symbols being determined
by the succesive iterations of this first value \cite{SanMartin07}.
\item [{Definition~2.}] The pattern $S_{n}$ has even {}``$R$-parity''
if it has an even number of symbols $R$ in it. Otherwise $S_{n}$
{}``$R$-parity'' will be termed odd \cite{MetropolisSteinStein73}.
\\
We will use $R$-parity whenever we work with logistic-like maps (upwards
convex). Should we were working with the real Mandelbrot map, for
instance (which is downwards convex), all $R$'s and $L$'s in what
follows would appear interchanged, and we would use $L$-parity instead.
\item [{Definition~3.}] The pattern $\overline{S_{n}}$ obtained exchanging
$R$ and $L$ in $S_{n}$ will be called the conjugated sequence of
$S_{n}$. In the case of a supercycle pattern $CS_{n-1}$ ($C$ corresponds
to $x_{1}=1/2$), its conjugate will only change the $R$ and $L$
in the pattern, whereas the initial $C$ remains in place.
\item [{Definition~4.}] $(S_{n})^{p}=\underbrace{S_{n}S_{n}\dots S_{n}}_{p\mbox{ times}}$,
is the pattern of $n\cdot p$ symbols constructed by $p$ repetitions
of the $n$-symbols pattern $S_{n}$. As a consequence, we may write
$\left[(S_{n})^{p}\right]^{q}=(S_{n})^{p\cdot q}$. We will adopt
an analogous notation for the case of supercycles.
\item [{Definition~5.}] $CS_{n-1}$ is a $n$-symbols pattern formed by
replacing the first letter in sequence $S_{n}$ by the symbol $C$.
This orbit corresponds to a supercycle periodic orbit (starting at
$x_{1}=\frac{1}{2}$).
\item [{Definition~6.}] $(CS_{n})^{p}\vert S_{p}$ is the pattern resulting
from the substitution of the $p$ succesive $C$-letters in the pattern
$(CS_{n})^{p}=\underbrace{CS_{n}CS_{n}\dots CS_{n}}_{p\mbox{ times}}$
by the $p$ succesive symbols in $S_{p}$.
\item [{Definition~7.}] $CP_{n,q}$ represents the pattern of the $n$-th
pitchfork bifurcation supercycle of the $q$-periodic orbit. For $q=1$
we will write simply $CP_{n}$.
\item [{Definition~8.}] The first F-harmonic of $CS_{n}$, denoted by
$H_{F}^{(1)}(CS_{n})$, is formed appending $CS_{n}$ to itself and
changing the second $C$ to $R$ ($L$) if the $R$-parity of $CS_{n}$
is even (odd). The second F-harmonic, $H_{F}^{(2)}(CS_{n})$, is formed
appending $CS_{n}$ to $H_{F}^{(1)}(CS_{n})$ and changing the new
$C$ to $R$ ($L$) is the $R$-parity of $H_{F}^{(1)}(CS_{n})$ is
even (odd). The successive F-harmonics are constructed in the same
way. \cite{Pastor96}

The F-harmonics (or Fourier harmonics) were introduced by Romera,
Pastor and Montoya in \cite{Pastor96} and have been widely employed
by these authors in the study of dynamical systems \cite{Romera96,Pastor02,Pastor03}.
It must be emphasized that, while those harmonics introduced by Metropolis,
Stein and Stein \cite{MetropolisSteinStein73} give us the patterns
corresponding to the Feigenbaum period doubling cascade, the F-harmonics
are used to compute the patterns of the last appearance orbits (see
Definition~10 below) in the current window. This property will be
amply used in our proofs; they will be, together with the saddle-node
bifurcation cascades, the main tools used in this paper.

\item [{Definition~9.}] A saddle-node bifurcation cascade is a sequence
of saddle-node bifurcations in which the number of fixed points showing
this kind of bifurcation is duplicated \cite{SanMartin07}. The successive
elements of the sequence are given by an equation identical to the
one that Feigenbaum found for a period doubling cascade \cite{Feigenbaum78,Feigenbaum79}.
This implies that both bifurcation cascades (Feigenbaum's and saddle-node)
scale in the same way.
\item [{Definition~10.}] We will consider a $q$-periodic window to be
a first appearance window within the $p$-periodic one if it corresponds
to values of parameter $\mu$ of expression (\ref{eq:logistic}) smaller
than any other $q$-periodic window inside the given $p$-periodic
one. Similarly, we will define a last appearance $q$-periodic orbit
(within the $p$-periodic one) as that having the larger $\mu$ parameter
among the period-$q$ windows (of the $p$-periodic one)\textbf{.}\\
A periodic window of period $q\cdot p$ will be called the $q$-periodic
window inside the given $p$-periodic one, to stress the similarity
of the structures in period-$p$ window with those in the canonical
window. 
\item [{Definition~11.}] Within the chaotic region of the logistic equation
bifurcation diagram (to the right of the Myrberg-Feigenbaum point),
there exist chaotic bands of $\mu$ values where the iterates of expression
(\ref{eq:logistic}) tend to be grouped in $2^{n}$ intervals (separated
by $2^{n}-1$ {}``empty'' intervals); we will refer to one of these
bands as the $n$-th chaotic band. Following the same naming convention
as in Definition~10, we will speak of the first and last appearance
$q\cdot2^{n}$-periodic orbits inside a chaotic band, as those corresponding
to the smaller and greater parameter $\mu$ values, respectively,
within that $n$-chaotic band.
\end{description}
In our proofs we will often use saddle-node cascades inside the canonical
window. They can be worked out applying the following theorem, which
is a particular case of that shown in \cite{SanMartin07}.

\begin{description}
\item [{Theorem~1}] (\emph{Saddle-Node bifurcation cascade} \emph{in the
canonical window}). \emph{The sequence of the $p\cdot2^{n}$-periodic
saddle-node orbit of a saddle-node bifurcation cascade starting from
a $p$-periodic orbit in the canonical window is obtained with the
following process:}

\begin{enumerate}
\item [(i)] \emph{Write the sequence of the orbit of the supercycle of
$f^{2^{n}}$, that is $CP_{n}$ (see Definition~7).}
\item [(ii)] \emph{Write consecutively $p$ times the sequence obtained
in the former point {}``i'', getting a sequence like\[
\underbrace{CP_{n}\dots CP_{n}}_{p\mbox{ times}}\]
}
\item [(iii)] \emph{Write the sequence of period-$p$ saddle-node orbit,
that is, the sequence of the saddle-node orbit of $f^{p}$. The first
point in the sequence must be the nearest saddle-node point to $C$.}
\item [(iv)] \emph{If $n$ is odd then conjugate the letters obtained in
point {}``iii'' by means of $L\leftrightarrow R$. Bear in mind
that $n\in{\cal Z}^{+}$.}
\item [(v)] \emph{Replace the $i$-th letter $C$ ($i=1,2,\dots p$) of
the sequence obtained in {}``ii'' by the $i$-th letter of the sequence
obtained in {}``iv''.}
\end{enumerate}
\end{description}

\section{\label{sec:CofC}Conjugation of cascades.}

\begin{description}
\item [{Theorem~2.}] (\emph{Conjugation of cascades} \emph{in the canonical
window})  \emph{Let $S_{p\cdot2^{n}}$ be the pattern of the $p\cdot2^{n}$-periodic
orbit of a saddle-node bifurcation cascade in the canonical window.
If $CS_{p\cdot2^{n}-1}$ denotes the supercycle associated to $S_{p\cdot2^{n}}$
(see Definition~5) and $CP_{n}$ denotes the supercycle of the $n$-th
pitchfork bifurcation (see Definition~7) then \[
CS_{p\cdot2^{n}-1}=H_{F}^{(p-1)}(CP_{n})\]
That is, the supercycle of the $n$-th element of the cascade can
be computed from the saddle-node bifurcation cascade (using Theorem~1)
or from the Feigenbaum cascade (using F-harmonics, see Definition~8).}
\end{description}
\textbf{Proof.} To prove this statement, we are going to compute the
pattern of one of these $2^{n}p$-periodic orbits in two ways: 1)
applying the \textbf{Theorem~1} \emph{(Saddle-Node bifurcation cascade
in the canonical window),} this way we get the $n$-th term of the
saddle-node bifurcation cascade, and 2) computing the $(p-1)$-th
F-harmonic to the $2^{n}$-periodic orbit generated after the $n$-th
pitchfork bifurcation of the $1$-periodic orbit.

In our proof we must consider separately the cases of even and odd
$n$.

\textbf{Let $n$ be even:}

\begin{enumerate}
\item [(i)] Applying Theorem~1:

Theorem~1 involves the following steps:

\begin{itemize}
\item determine the last apperarance $p$-periodic orbit pattern, that is
$CRL^{p-2}$ \cite{Post91}. Given that $n$ is even, this pattern
will not be conjugated (i.e. $R$'s and $L$'s will not be swapped).
\item repeat $p$ times the sequence $CP_{n}$ which gives\begin{equation}
\underbrace{CP_{n}CP_{n}\dots CP_{n}}_{p\mbox{ times}}\label{eq:th1patt1}\end{equation}

\item substitute the succesive $C$ symbols in the pattern (\ref{eq:th1patt1})
by the succesive symbols in the pattern $CRL^{p-2}$, which gives
the pattern sought for:\begin{equation}
CP_{n}RP_{n}(LP_{n})^{p-2}\label{eq:th1patt2}\end{equation}
where $p=3,\,4,\,5,\dots$ (there is no cascade for $p=2$).
\end{itemize}
\item [(ii)] Applying F-harmonics:

The construction of the F-harmonics follows the next steps:

\begin{itemize}
\item append $CP_{n}$ to the pattern $CP_{n}$, giving\[
CP_{n}CP_{n}\]

\item substitute the second $C$ by $R$ (or $L$) if the $R$-parity of
$CP_{n}$ is even (or odd). Given that $n$ is assumed to be even,
the $R$-parity of $CP_{n}$ is also even (see Note~1 in Appendix);
as a consequence the first F-harmonic becomes\[
H_{F}^{(1)}(CP_{n})=CP_{n}RP_{n}\]

\item generate the $n$-th F-harmonic appending $CP_{n}$ to the $(n-1)$-th
F-harmonic, and changing the second $C$ of the obtained sequence
by $R$ (or $L$) if the $R$-parity of $H_{F}^{(n-1)}(CP_{n})$ is
even (or odd). It is easy to see that $H_{F}^{(1)}(CP_{n})$ $R$-parity
is odd. As a consequence, on appending the expression $CP_{n}$ to
$H_{F}^{(1)}(CP_{n})$ in order to obtain the $H_{F}^{(2)}(CP_{n})$,
the $C$ will be changed to $L$, thus conserving $R$-parity. The
same will happen for the next F-harmonics thus getting\begin{equation}
H_{F}^{(p-1)}(CP_{n})=CP_{n}RP_{n}(LP_{n})^{p-2}\label{eq:th1patt3}\end{equation}
where $p=2,\,3,\,4,\dots$ (although in the cascade there is no $p=2$).
\end{itemize}
\end{enumerate}
~

\textbf{Let $n$ be odd:}

\begin{enumerate}
\item [(iii)] Applying Theorem~1:

In this case of $n$ odd, Theorem~1 would be equally applied but
pattern $CRL^{p-2}$ must be conjugated to $CLR^{p-2}$. Thus the
final pattern will be\begin{equation}
CP_{n}LP_{n}(RP_{n})^{p-2}\label{eq:th1patt4}\end{equation}

\item [(iv)] Applying F-harmonics:

Being $n$ odd, the $R$-parity of $CP_{n}$ is also odd (see Note~1
in Appendix), thus modifying the first F-harmonic to\[
H_{F}^{(1)}(CP_{n})=CP_{n}LP_{n}\]
having even $R$-parity. As a consequence, on appending $CP_{n}$
to $H_{F}^{(1)}(CP_{n})$ to get to $H_{F}^{(2)}(CP_{n})$ the second
$C$ will be replaced by $R$ thus conserving the $R$-parity. The
result is\[
H_{F}^{(p-1)}(CP_{n})=CP_{n}LP_{n}(RP_{n})^{p-2}\]
with $p=2,\,3,\,4,\,5,\dots$

\end{enumerate}
~

In summary, the theorem is proven because both Theorem~1 and the
F-harmonics give the same results \begin{equation}
\begin{array}{ll}
CP_{n}RP_{n}(LP_{n})^{p-2} & \, p=3,4,5,\dots\mbox{ for }n\mbox{ even}\\
CP_{n}LP_{n}(RP_{n})^{p-2} & \, p=3,4,5,\dots\mbox{ for }n\mbox{ odd}\end{array}\label{eq:th1canonic}\end{equation}
The terms with $p=2$ in the F-harmonic have been disregarded, given
that they correspond to a pitchfork bifurcation which, thus, cannot
account for a saddle-node one.$\blacktriangleleft$

\subsection*{Examples.}

As an example, we will show how to obtain the first saddle-node bifurcation
orbits of the last appearance period-$5$ orbit in the canonical window,
whose pattern is $CRL^{3}$. It will be enough to examine the cases
$n=1,2$ to show how to perform the calculation in both cases when
the $P_{n}$ orbit has to be conjugated or not.

For $n=1$, that is odd $n$:

\begin{itemize}
\item Using Theorem~1 we compute:

\begin{itemize}
\item the corresponding $CP_{1}$ in the canonical window is simply $CR$
\item the pattern $CRL^{3}$ will be conjugated because $n$ is odd, resulting:
$CRL^{3}\rightarrow CLR^{3}$
\item repeat $CR$ $p=5$ times: $CRCRCRCRCR$
\item replace $C$'s in the latter sequence by $CLR^{3}$, resulting: $CRLR^{7}$
\end{itemize}
\item Using Theorem~2 and F-harmonics:

\begin{itemize}
\item $CP_{1}=CR$
\item its first F-harmonic is $CRCR\rightarrow CRLR$ ($C$ changes to $L$
because $R$-parity of $CR$ is odd)
\item its second F-harmonic is $CRLRCR\rightarrow CRLR^{3}$ ($C$ changes
to $R$ because $R$-parity of $CRLR$ is even)
\item and its fourth F-harmonic is $CRLR^{7}$
\end{itemize}
\end{itemize}
Both sequences coincide, as it was expected.

For $n=2$, that is even $n$:

\begin{itemize}
\item Using Theorem~1 we compute:

\begin{itemize}
\item $CP_{2}=CRLR$
\item the pattern $CRL^{3}$ will not be conjugated because $n$ is even
\item repeat $CRLR$ $p=5$ times: $CRLRCRLRCRLRCRLRCRLR$
\item replace $C$'s in the latter sequence by $CRL^{3}$, resulting: $CRLR^{3}(LR)^{7}$
\end{itemize}
\item Using Theorem~2 and F-harmonics:

\begin{itemize}
\item $CP_{2}=CRLR$
\item its first F-harmonic is $CRLRCRLR\rightarrow CRLR^{3}LR$ ($C$ changes
to $R$ because $R$-parity of $CRLR$ is even)
\item and its fourth F-harmonic is $CRLR^{3}(LR)^{7}$
\end{itemize}
\end{itemize}
Both sequences also coincide.

\section{\label{sec:1st-and-last}Convergence of first and last appearance
orbits.}

\begin{description}
\item [{Theorem~3.}] \emph{In the canonical window, the last appearance
orbits within the $(n+1)$-chaotic band have the same accumulation
point as the first appearance orbits within the $n$-chaotic band.}
\end{description}
\textbf{Proof.} To prove the theorem, we need to compute the patterns
of the first and last appearance orbits within an arbitrary chaotic
band. These patterns are given by the saddle-node bifurcation cascades
associated, respectively, to the first and last orbits in the $1$-chaotic
band. The saddle-node cascades will be computed by means of Theorem~1.
Given that this theorem treats the cases of even and odd $n$ differently,
we will study these two cases separately.

\textbf{Let $n$ be even:}

\begin{enumerate}
\item [(i)] Computation of the first appearance orbit patterns within the
$n$-chaotic band.

As we have already indicated, we compute the requested pattern applying
Theorem~1 following the next steps:

\begin{itemize}
\item give the pattern of the first appearance superstable $p$-periodic
orbit within the $1$-chaotic band. This pattern is $CRLR^{p-3}$,
with $p=3,5,7,\dots$ 
\item repeat $p$ times the pattern $CP_{n}$ (see Definition~7) as in
the following\begin{equation}
\underbrace{CP_{n}CP_{n}\dots CP_{n}}_{p\mbox{ times}}\label{eq:th2patt1}\end{equation}
Given that we assume $n$ to be even, the symbols in the pattern $CRLR^{p-3}$
are not conjugated, and succesive $C$'s in expression (\ref{eq:th2patt1})
are simply replaced by the successive $C$'s in the pattern $CRLR^{p-3}$.
This substitution results in the sought for pattern\begin{equation}
CP_{n}RP_{n}LP_{n}(RP_{n})^{p-3}\label{eq:th2patt2}\end{equation}
where $p=3,\,5,\,7,\dots$ As expected, the pattern (\ref{eq:th2patt2})
has $p\cdot2^{n}$ symbols.
\end{itemize}
\item [(ii)] Computation of the last appearance orbit patterns within the
$(n+1)$-chaotic band.

To apply Theorem~1 we accomplish the following steps:

\begin{itemize}
\item give the pattern of the last appearance superstable $p$-periodic
orbit within the $1$-chaotic band, that is \cite{Post91}\[
CRL^{p-2}\]
with $p=3,\,4,\,5,\dots$ 
\item repeat $p$ times the pattern $CP_{n+1}$. Given that $n$ is even,
$CP_{n+1}=CP_{n}RP_{n}$ (see Note~1 in Appendix), this becomes\begin{equation}
\underbrace{CP_{n}RP_{n}CP_{n}RP_{n}\dots CP_{n}RP_{n}}_{p\mbox{ times}}\label{eq:th2patt3}\end{equation}

\item given that $n+1$ is odd the pattern $CRL^{p-2}$ will be conjugated,
resulting in $CLR^{p-2}$. Next, succesive $C$'s in expression (\ref{eq:th2patt3})
are replaced by successive symbols in $CLR^{p-2}$. This substitution
gives the sought for pattern, that is\[
CP_{n}RP_{n}LP_{n}(RP_{n})^{2p-3}\]
where $p=3,\,4,\,5,\dots$
\end{itemize}
\end{enumerate}
~

\textbf{Let $n$ be odd:}

\begin{enumerate}
\item [(iii)] Computation of the first appearance orbit patterns within
the $n$-chaotic orbit.

We take an identical approach to {}``i'', but given that $n$ is
odd, the pattern $CRLR^{p-3}$ must be conjugated to $CLRL^{p-3}$.
The consequence of this conjugation is that the following pattern
is generated\[
CP_{n}LP_{n}RP_{n}(LP_{n})^{p-3}\]
with $p=3,\,5,\,7,\dots$

\item [(iv)] Computation of the last appearance orbit patterns within the
$(n+1)$-chaotic orbit.

We take an identical approach to {}``ii'', but given that $(n+1)$
is even, the pattern $CRL^{p-2}$ must not be conjugated. On the other
hand, given that $n$ is odd, we have $CP_{n+1}=CP_{n}LP_{n}$ (see
Note~1 in Appendix). As a consequence, the pattern \[
CP_{n}LP_{n}RP_{n}(LP_{n})^{2p-3}\]
is obtained, with $p=3,\,4,\,5,\dots$

\end{enumerate}
~

In summary we get:

\begin{itemize}
\item the first appearance orbit patterns within the $n$-chaotic band are:
\begin{equation}
\begin{array}{ll}
CP_{n}RP_{n}LP_{n}(RP_{n})^{p-3} & \, p=3,5,7,\dots\mbox{ for }n\mbox{ even}\\
CP_{n}LP_{n}RP_{n}(LP_{n})^{p-3} & \, p=3,5,7,\dots\mbox{ for }n\mbox{ odd}\end{array}\label{eq:bandas-canonica_1}\end{equation}

\item the last appearance orbit patterns within the $(n+1)$-chaotic band
are:\begin{equation}
\begin{array}{ll}
CP_{n}RP_{n}LP_{n}(RP_{n})^{2p-3} & \, p=3,4,5,\dots\mbox{ for }n\mbox{ even}\\
CP_{n}LP_{n}RP_{n}(LP_{n})^{2p-3} & \, p=3,4,5,\dots\mbox{ for }n\mbox{ odd}\end{array}\label{eq:bandas-canonica_2}\end{equation}

\end{itemize}
Myrberg's formula \cite{Myrberg63} shows that the parameters corresponding
to these orbits have the same accumulation point, and this proves
the theorem. $\blacktriangleleft$

\begin{figure}[tph]
\begin{centering}\includegraphics[width=1\textwidth]{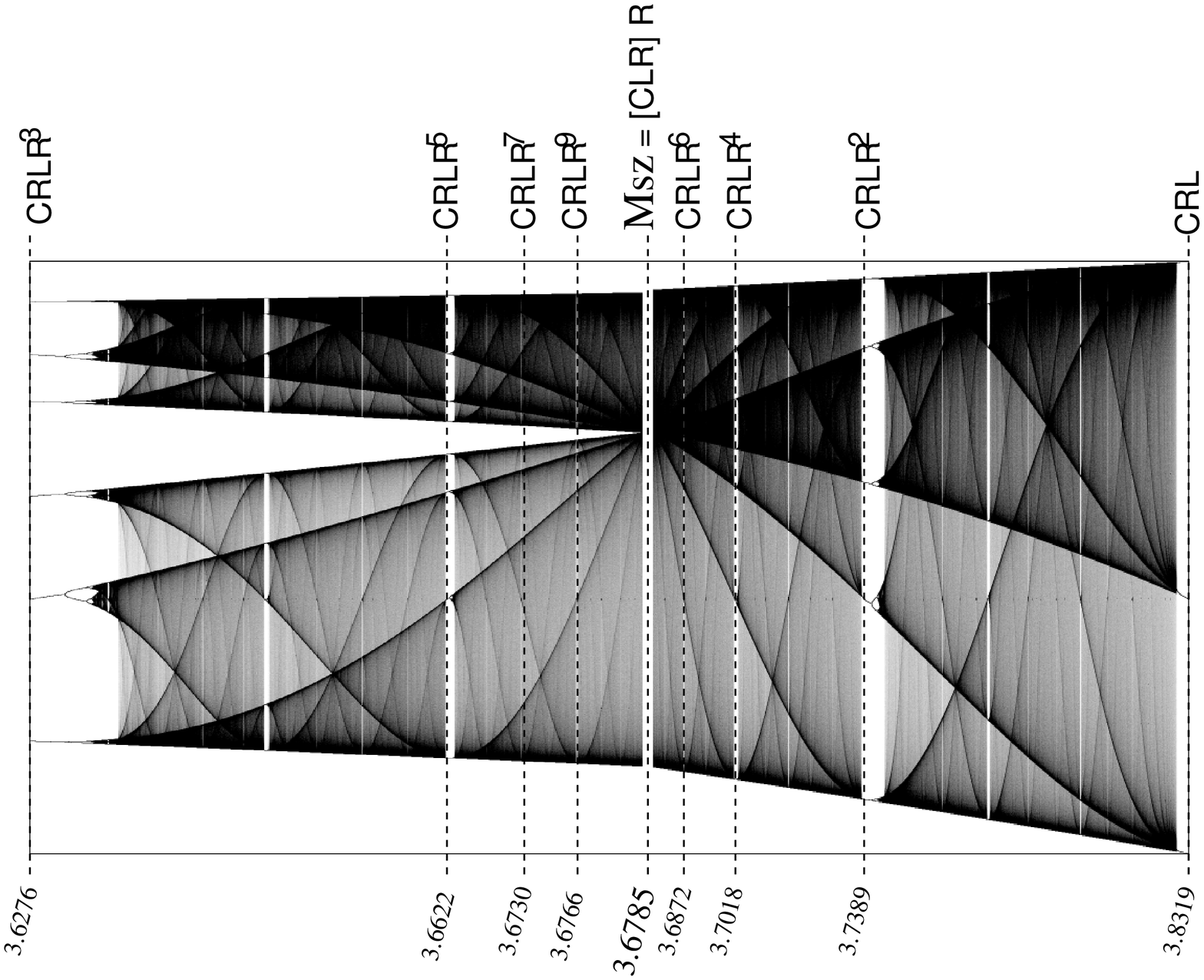}\par\end{centering}

\caption{\label{fig:FandL-app-orbits}First and last appearance orbits approaching
the band merging Misiurewicz point.}
\end{figure}

\subsection{Consequences of Theorem~3 (Misiurewicz point cascade)}

The limit $p\rightarrow\infty$ of expressions (\ref{eq:bandas-canonica_1})
and (\ref{eq:bandas-canonica_2}) corresponds to those points where
the $n$ and $n+1$ bands merge. These are Misiurewicz points having
a preperiod (within square brackets in the following expression) followed
by a period that corresponds to an unstable periodic orbit:\begin{equation}
\begin{array}{rl}
{}[CP_{n}RP_{n}LP_{n}]\; RP_{n} & \quad\mbox{ for }n\mbox{ even}\\
{}[CP_{n}LP_{n}RP_{n}]\; LP_{n} & \quad\mbox{ for }n\mbox{ odd}\end{array}\label{eq:misiurewicz-A}\end{equation}
These expressions were previously obtained in \cite{Romera96} using
a different approach. The value of the parameter $\mu$ in expression
(\ref{eq:logistic}), where these points are localized, is the smallest
solution of the equation \[
{\cal P}_{2^{n}+1}(\mu)={\cal P}_{2^{n-1}}(\mu)\]
being ${\cal P}_{n}(\mu)$ the dynamical polynomials defined in \cite{Myrberg63,Zheng84},
which can be obtained recursively (for equation (\ref{eq:logistic}))
as\[
{\cal P}_{0}=1/2,\quad{\cal P}_{n+1}=\mu{\cal P}_{n}(1-{\cal P}_{n})\]

Given that patterns (\ref{eq:misiurewicz-A}) have been obtained as
the limit of saddle-node orbit sequences elements
(each of these elements belongs to a different saddle-node bifurcation
cascade) the following corollary can be stated:

\begin{description}
\item [{Corollary~to~Theorem~3.}] \emph{Band merging points, whose patterns
are given in (\ref{eq:misiurewicz-A}) as the limit orbits of saddle-node
orbit sequences, form a Misiurewicz point cascade. Furthermore, this
Misiurewicz point cascade converges to the Myrberg-Feigenbaum point,
driven by the behavior of the saddle-node bifurcation cascades, as
is described in \cite{SanMartin07}.}
\end{description}

\section{\label{sec:generalization}Generalization to any $q$-periodic window}

The selfsimilarity shown by the behaviors of the logistic equation
is well known. It is revealed by the existence of periodic windows
that mimic the different structures of the canonic window. Inside
every subwindow the same bifurcation diagram structure exists, with
its periodic and chaotic regions, intermittencies, and so on. In particular,
every window shows in its beginning a periodic behavior, followed
by a pitchfork bifurcation cascade; similarly every window has saddle-node
bifurcation cascades, which reflect the ones existing in the canonical
window.

This circumstance is a clear suggestion to look for the relationship
between the orbits of the pitchfork bifurcation cascade and those
of the saddle-node bifurcation cascade, both belonging to a given
$q$-periodic window. In this way we generalize the results obtained
above, for the canonical window, to any arbitrary window. Nonetheless,
although the results for the canonical window will become a particular
case of the more general theorems, they will be used in the proofs
of the general theorems.

The Bifurcation Rigidity theorem \cite{Hunt99} allows us to approximately
compute the parameter values for which the different orbits in a $p$-periodic
window, mimicking those in the canonical one, take place. This is
achieved through a linear mapping of the windows, the deeper the subwindow
is located the more exact the result is. Nevertheless, this useful
tool does not allow us to determine the patterns of the orbits and
thus, we cannot establish a relationship among them.

To obtain the patterns of the orbits, we need to compute that of the
$p\cdot2^{n}$-periodic orbits in a saddle-node bifurcation cascade,
located inside a $q$-periodic window. A similar problem (that of
finding the pattern of a $q$-periodic orbit inside the $p\cdot2^{n}$-periodic
window) was already addressed in \cite{SanMartin07} and is thus known
to us. Merely exchanging the names of the windows involved, brings
us the requested pattern. This result, using the compact notation
introduced in section \ref{sec:Definitions}, is stated as follows.

\begin{description}
\item [{Theorem~4.}] \emph{Let $S_{p\cdot2^{n}}$ ($n=1,2,3\dots$) be
the patterns of the orbits in a saddle-node bifurcation cascade located
inside the canonical window. Let $S_{q}$ be the pattern of the $q$-periodic
saddle-node orbit which marks the origin of the $q$-periodic window,
and let $CS_{q-1}$ be the pattern of its associated supercycle. The
pattern $S_{q\cdot p\cdot2^{n}}$ of the $p\cdot2^{n}$-periodic orbit
in the saddle-node bifurcation cascade inside the $q$-periodic window
considered is \[
\left((CS_{q-1})^{p\cdot2^{n}}\vert S_{p\cdot2^{n}}\right)\qquad\mbox{for even }R\mbox{-parity of the }q\mbox{-periodic supercycle}\]
\[
\left((CS_{q-1})^{p\cdot2^{n}}\vert\overline{S_{p\cdot2^{n}}}\right)\qquad\mbox{for odd }R\mbox{-parity of the }q\mbox{-periodic supercycle}\]
}
\end{description}
\textbf{Proof.} The proof is achieved by just exchanging $f^{q}\leftrightarrow f^{p\cdot2^{n}}$
in the original proof (see \cite{SanMartin07}). With this substitution,
expression \[
\left((CS_{q-1})^{p\cdot2^{n}}\vert S_{p\cdot2^{n}}\right)\]
is obtained, which is valid for $f^{q}$ having a maximum at $x=1/2$.
This corresponds to an even $R$-parity of the $q$-periodic orbit
(see Lemma~1 in the appendix). Should $f^{q}$ have a minimum at
$x=1/2$, that is, for a $q$-periodic orbit having odd $R$-parity
(see Lemma~2 in the appendix), it is enough to conjugate $S_{p\cdot2^{n}}\rightarrow\overline{S_{p\cdot2^{n}}}$
to get the correct result, that is\[
\left((CS_{q-1})^{p\cdot2^{n}}\vert\overline{S_{p\cdot2^{n}}}\right)\]

$\blacktriangleleft$

\begin{figure}[tph]
\begin{centering}\includegraphics[width=0.7\textwidth]{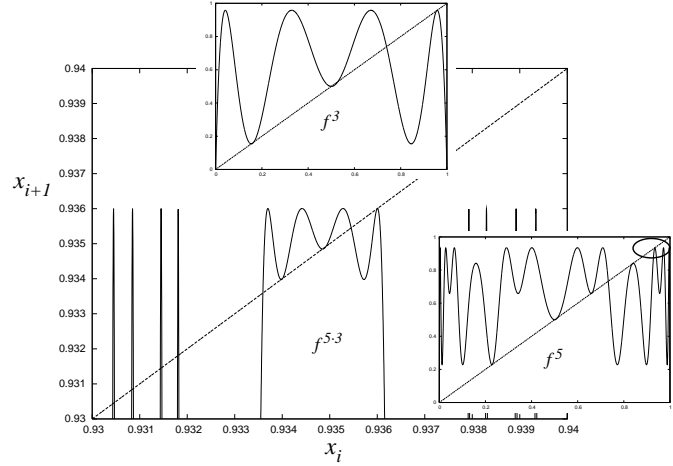}\par\end{centering}

\caption{\label{fig:3in5}Graph of $f^{5\cdot3}$ (the $f^{3}$ periodic orbit
inside the period-$5$ window). It can be seen how in one of the extrema
of $f^{5}$ (highlighted in the right panel plot) the shape of $f^{3}$
is reproduced.}
\end{figure}

\begin{figure}[tph]
\begin{centering}\includegraphics[width=0.6\textwidth]{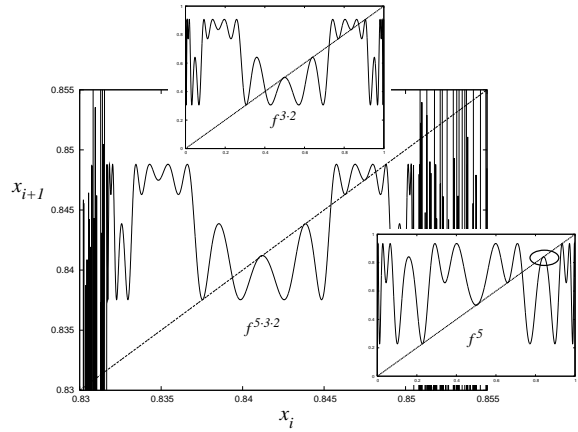}\par\end{centering}

\caption{\label{fig:3x2in5}Graph of $f^{5\cdot3\cdot2}$ (the $f^{3\cdot2}$
saddle-node bifurcation of the $f^{3}$ periodic orbit inside the
period-$5$ window, which is located in the $1$-chaotic band of the
period-$5$ window). It can be seen how in one of the extrema of $f^{5}$
(drawn in the right panel and encircled by the ellipse) the shape
of $f^{3\cdot2}$ is reproduced.}
\end{figure}

As a consequence of replacing $f^{q}\leftrightarrow f^{p\cdot2^{n}}$,
we observe that the orbit of $f^{p\cdot2^{n}}$ gets reproduced around
every extremum of $f^{q}$ lying near the line $x_{i+1}=x_{i}$. This
is shown in figure \ref{fig:3in5}, where the shape of the $f^{3}$
orbit (upper panel in figure \ref{fig:3in5}) can be identified in
those extremes of $f^{5}$ (right panel in figure \ref{fig:3in5})
near the diagonal $x_{i+1}=x_{i}$. In figure \ref{fig:3x2in5} the
same situation is shown, this time for the supercycle of $f^{5\cdot3\cdot2}$
(saddle-node bifurcation of that of $f^{5\cdot3}$); in this case
it is $f^{3\cdot2}$ that gets reproduced in each maximum of $f^{5}$.

Let us emphasize that, what is being obtained is the pattern of a
$p\cdot2^{n}$-periodic saddle-node orbit located inside a $q$-periodic
window. Given that this $q$-periodic window has its origin at a $q$-periodic
saddle-node orbit, the resulting pattern will be that of a $q\cdot p\cdot2^{n}$-periodic
saddle-node orbit. Moreover, the geometrical construction used in
the proof, allows us to show the geometrical meaning of the DGP formula,
as can be seen in what follows.

\begin{figure}[tph]
\begin{centering}\includegraphics[width=0.8\textwidth]{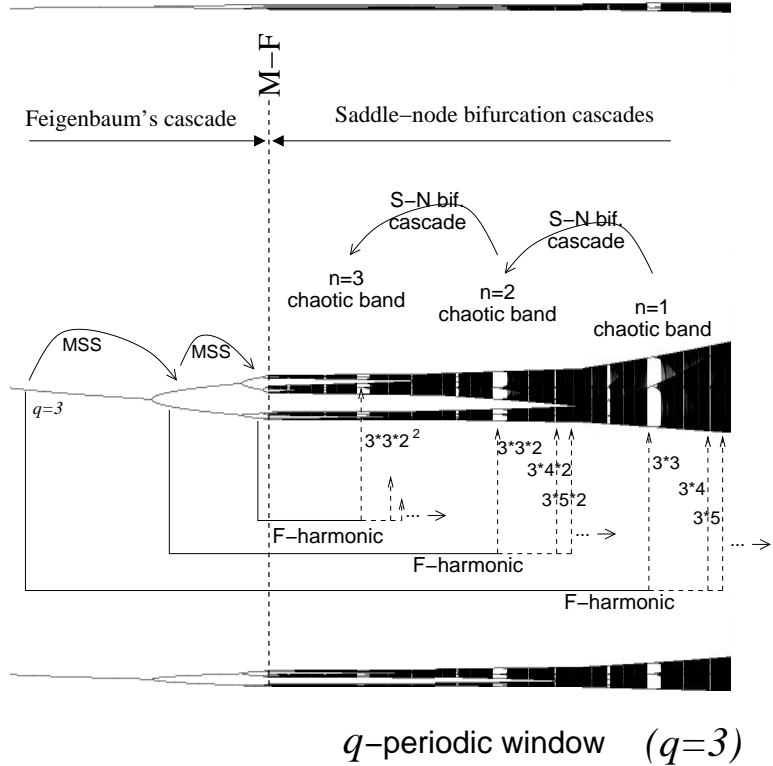}\par\end{centering}

\caption{\label{fig:p2n-in-e}Example of the application of Theorem~5, within
the period-$3$ window to find the last appearance orbits in the $1$-
and $2$-chaotic bands, applying F-harmonics to the orbits in the
Feigenbaum cascade. Also shown, is how the Feigenbaum cascade is computed
from Metropolis-Stein-Stein harmonics \cite{MetropolisSteinStein73},
and the saddle-node bifurcation cascade is obtained using Theorem~4.}
\end{figure}

The substitution of the first symbol in the pattern $S_{q\cdot p\cdot2^{n}}$
by a $C$ is enough to obtain its associated supercycle, that we denote
by $CS_{q\cdot p\cdot2^{n}-1}$ (see Definition~5), whose geometrical
interpretation was given above. Given that this same supercycle can
be also constructed applying DGP rule to compute the product of the
orbits $CS_{p\cdot2^{n}-1}$ and $CS_{q-1}$, what is obtained is
a geometrical interpretation of the DGP rule. We thus conclude (see
figure \ref{fig:p2n-in-e} for a graphical sketch) that:

\begin{enumerate}
\item The composition of the supercycles $CS_{p\cdot2^{n}-1}$ and $CS_{q-1}$
, according to DGP rule, represents the supercycle associated to the
$p\cdot2^{n}$-periodic orbit, inside the $q$-periodic window.

That is so because, in the geometric proof, we have forced the shape
of a $p\cdot2^{n}$-periodic saddle-node orbit to appear around the
extrema of a saddle-node $q$-periodic orbit. This is shown in figures
\ref{fig:3in5} and \ref{fig:3x2in5}.

\item However, we could have forced the reverse situation, that is, to reproduce
the shape of the $q$-periodic orbit around the extrema of a $p\cdot2^{n}$-periodic
orbit, arising after $n$ pitchfork bifurcations of a $p$-periodic
saddle-node orbit. In that case, we would be focusing on the $n$-th
pitchfork bifurcation of a $p$-periodic orbit, localized inside the
$q$-periodic window. This second case was already known and is not
a conclusion of our geometrical proof.
\end{enumerate}

Theorem~4 allows us to state Theorem~2 and Theorem~3 generalized
to subwindows inside the canonical one. The proofs are given in the
next two subsections.

\subsection{\label{sub:CoC-general}Conjugation of cascades in an arbitrary window.}

\begin{description}
\item [{Theorem~5}] (\emph{Conjugation of cascades in an arbitrary window}).
\emph{Let $S_{q\cdot p\cdot2^{n}}$ be the pattern of the $p\cdot2^{n}$-periodic
orbit in the saddle-node bifurcation cascade inside the $q$-periodic
window, such that $p$ is a last appearance orbit. If $CS_{q\cdot p\cdot2^{n}-1}$
denotes the supercycle associated to $S_{q\cdot p\cdot2^{n}}$ (see
Definition~5) and $CP_{n,q}$ denotes the $n$-th pitchfork bifurcation
supercycle of the $q$-periodic orbit (see Definition~7) then\[
CS_{q\cdot p\cdot2^{n}-1}=H_{F}^{(p-1)}(CP_{n,q})\]
}

What the theorem says is that, inside an arbitrary $q$-periodic window,
superstable orbit patterns of a saddle-node bifurcation cascade, which
is associated to a last appearance orbit, coincide with orbit patterns
generated by appying F-harmonics to Feigenbaum cascade supercycles.

\end{description}
\textbf{Proof.} Let us consider a $q$-periodic window, whose origin
is in a $q$-periodic saddle-node orbit. This $q$-periodic window
needs not to be a primary window, any window located inside any other
will do.

Let us also consider the last appearance $p$-periodic orbit inside
the chosen $q$-periodic window, where $p=3,\,4,\,5,\dots$. This
$p$-periodic orbit has a saddle-node bifurcation cascade associated
to it, whose orbits have periods $p\cdot2^{n}$, $n=0,\,1,\,2,\dots$,
and are all located inside the $q$-periodic window (that is, they
are really $q\cdot p\cdot2^{n}$-periodic orbits).

We want to show that the $p\cdot2^{n}$-periodic saddle-node orbit
supercycle inside the $q$-periodic window can be obtained equivalently
by applying Theorem~4 or computing the $(p-1)$-th F-harmonic of
the $n$-th pitchfork bifurcation of the $q$-periodic orbit.

~

\textbf{Let the $R$-parity of the $q$-periodic supercycle pattern
be even.}

\begin{enumerate}
\item [(i)]Applying Theorem~4.

This theorem (see section \ref{sec:generalization}) says that the
pattern $CS_{q\cdot p\cdot2^{n}-1}$ is given by\[
\left((CS_{q-1})^{p\cdot2^{n}}\vert CS_{p\cdot2^{n}-1}\right)\mbox{ for even }R\mbox{-parity of the }q\mbox{-periodic orbit}\]
On the other hand, the pattern $CS_{p\cdot2^{n}-1}$ is given by expression
(\ref{eq:th1canonic}).

\begin{enumerate}
\item Let $n$ be odd. Then $CS_{p\cdot2^{n}-1}=CP_{n}LP_{n}(RP_{n})^{p-2}$
according to expression (\ref{eq:th1canonic}). Then we perform the
following computation\[
\begin{array}{rl}
\left((CS_{q-1})^{p\cdot2^{n}}\vert CS_{p\cdot2^{n}-1}\right)= & \left((CS_{q-1})^{p\cdot2^{n}}\vert CP_{n}LP_{n}(RP_{n})^{p-2}\right)\\
= & \left((CS_{q-1})^{2^{n}}\vert CP_{n}\right)\left((CS_{q-1})^{2^{n}}\vert LP_{n}\right)\\
 & \left((CS_{q-1})^{2^{n}}\vert RP_{n}\right)^{p-2}\end{array}\]

\item Let $n$ be even. Then $CS_{p\cdot2^{n}-1}=CP_{n}RP_{n}(LP_{n})^{p-2}$,
with $p=3,\,4,\,5,\dots$ according to expression (\ref{eq:th1canonic}).
In this case the computation goes as follows\[
\begin{array}{rl}
\left((CS_{q-1})^{p\cdot2^{n}}\vert CS_{p\cdot2^{n}-1}\right)= & \left((CS_{q-1})^{p\cdot2^{n}}\vert CP_{n}RP_{n}(LP_{n})^{p-2}\right)\\
= & \left((CS_{q-1})^{2^{n}}\vert CP_{n}\right)\left((CS_{q-1})^{2^{n}}\vert RP_{n}\right)\\
 & \left((CS_{q-1})^{2^{n}}\vert LP_{n}\right)^{p-2}\end{array}\]

\end{enumerate}
\item [(ii)]Applying F-harmonics.

Let us now apply F-harmonics to the $n$-th pitchfork bifurcation
of the $q$-periodic orbit. 

The $n$-th pitchfork bifurcation of $CS_{q-1}$ , that is $CP_{n,q}$,
is given by \[
\left((CS_{q-1})^{2^{n}}\vert CP_{n}\right)\]
and has the same $R$-parity as $CP_{n}$, according to Lemma~2 in
the Appendix.

\begin{enumerate}
\item Let $n$ be odd, then the $R$-parity of $CP_{n}$ is odd and, hence,
the $R$-parity of $\left((CS_{q-1})^{2^{n}}\vert CP_{n}\right)$
is also odd (for $n\geq1$). The F-harmonic is computed as\[
\begin{array}{l}
H_{F}^{(1)}\left[(CS_{q-1})^{2^{n}}\vert CP_{n}\right]=\\
=\left[(CS_{q-1})^{2^{n}}\vert CP_{n}\right]\left[(CS_{q-1})^{2^{n}}\vert LP_{n}\right]\end{array}\]
where the $C$'s in the last bracket have been changed to $L$'s because
of the assumed odd $R$-parity. As a consequence, $H_{F}^{(1)}\left[(CS_{q-1})^{2^{n}}\vert CP_{n}\right]$
has even $R$-parity. To compute the second F-harmonic $H_{F}^{(2)}\left[(CS_{q-1})^{2^{n}}\vert CP_{n}\right]$,
those $C$'s will be replaced by $R$ thus keeping the same $R$-parity.
Hence, the result for the $(p-1)$-th F-harmonic is \[
\begin{array}{l}
H_{F}^{(p-1)}\left[(CS_{q-1})^{2^{n}}\vert CP_{n}\right]=\\
=\left[(CS_{q-1})^{2^{n}}\vert CP_{n}\right]\left[(CS_{q-1})^{2^{n}}\vert LP_{n}\right]\left[(CS_{q-1})^{2^{n}}\vert RP_{n}\right]^{p-2}\end{array}\]

\item Let $n$ be even, thus the $R$-parity of $CP_{n}$ is even as well
and the $R$-parity of $\left((CS_{q-1})^{2^{n}}\vert CP_{n}\right)$
is also even. The first F-harmonic of $\left((CS_{q-1})^{2^{n}}\vert CP_{n}\right)$
is\[
H_{F}^{(1)}\left[(CS_{q-1})^{2^{n}}\vert CP_{n}\right]=\left[(CS_{q-1})^{2^{n}}\vert CP_{n}\right]\left[(CS_{q-1})^{2^{n}}\vert RP_{n}\right]\]
It is easy to see that $H_{F}^{(1)}\left[(CS_{q-1})^{2^{n}}\vert CP_{n}\right]$
has odd $R$-parity. As a consequence, on appending the expression
$\left[(CS_{q-1})^{2^{n}}\vert CP_{n}\right]$ to $H_{F}^{(1)}\left[(CS_{q-1})^{2^{n}}\vert CP_{n}\right]$,
in order to obtain the second F-harmonic, the $C$ will be switched
to $L$, keeping the same $R$-parity. The same will happen for the
next F-harmonics, thus giving\[
\begin{array}{rl}
H_{F}^{(p-1)}\left[(CS_{q-1})^{2^{n}}\vert CP_{n}\right]= & \left[(CS_{q-1})^{2^{n}}\vert CP_{n}\right]\left[(CS_{q-1})^{2^{n}}\vert RP_{n}\right]\\
 & \left[(CS_{q-1})^{2^{n}}\vert LP_{n}\right]^{p-2}\end{array}\]

\end{enumerate}
\end{enumerate}
We have thus proven that, for even as well as for odd $n$, the same
results are achieved both ways.

\textbf{Let the $R$-parity of the $q$-periodic supercycle pattern
be odd.} In this case we will have to compute \[
\left((CS_{q-1})^{p\cdot2^{n}}\vert\overline{CS_{p\cdot2^{n}-1}}\right)\]
using Theorem~4. On the other hand, the $n$-th pitchfork bifurcation
will be \[
\left((CS_{q-1})^{2^{n}}\vert\overline{CP_{n}}\right)\]
and it has the same $R$-parity as $\overline{CP_{n}}$ according
to Lemma~2 in the Appendix. Furthermore $\overline{CP_{n}}$ has
odd (even) $R$-parity when $n$ is even (odd) as indicated in Note~1.
Therefore, when F-harmonics are calculated, the same results will
be obtained.$\blacktriangleleft$

\subsection{\label{sub:1st-and-last-general}Convergence of first and last appearance
orbits in an arbitrary window.}

\begin{description}
\item [{Theorem~6:}] \emph{Inside any $q$-periodic window, the last apperance
orbits within the $(n+1)$-chaotic band have the same accumulation
point as the first appearance orbits within the $n$-chaotic band.}
\end{description}
\textbf{Proof.} Once again we study separately the
cases of even and odd $R$-parity:

\textbf{Let the $R$-parity of the $q$-periodic supercycle pattern
be even}. We want to compute the pattern of the supercycles of the
$p\cdot2^{n}$ first appearance and $p\cdot2^{n+1}$ last appearance
saddle-node orbits within the $q$-window. These orbits are located
in the $n$- and $(n+1)$-chaotic bands of that window, respectively.
In order to do so, we will apply Theorem~4, and hence we need to
know the patterns of the $p\cdot2^{n}$ first appearance and $p\cdot2^{n+1}$
last appearance saddle-node orbits within the canonical window, which
are located in the $n$- and $(n+1)$-chaotic bands, respectively.

\begin{enumerate}
\item [(i)]Let $n$ be even.

The pattern of the supercycle associated to the first appearance $p\cdot2^{n}$
saddle-node orbit, in the canonical window, is (see \ref{eq:bandas-canonica_1})
\begin{equation}
CS_{p\cdot2^{n}-1}=CP_{n}RP_{n}LP_{n}(RP_{n})^{p-3}\label{eq:th2Gpatt1}\end{equation}
where $p=3,\,5,\,7,\,\dots$. Also, the pattern of the supercycle
of the last appearance $p\cdot2^{n+1}$ saddle-node orbit, in the
canonical window, is given by (see \ref{eq:bandas-canonica_2})\begin{equation}
CS_{p\cdot2^{n+1}-1}=CP_{n}RP_{n}LP_{n}(RP_{n})^{2\cdot p-3}\label{eq:th2Gpatt2}\end{equation}

\begin{itemize}
\item The pattern of the supercycle associated to the $p\cdot2^{n}$ first
appearance saddle-node orbit, within the $q$-window, is computed
by applying Theorem~4 as follows:

\begin{enumerate}
\item Repeat $p\cdot2^{n}$ times the pattern $CS_{q-1}$ \[
(CS_{q-1})^{p\cdot2^{n}}\]

\item Conjugate it according to (\ref{eq:th2Gpatt1})\begin{equation}
\begin{array}{r}
\left[(CS_{q-1})^{2^{n}}\vert CP_{n}\right]\left[(CS_{q-1})^{2^{n}}\vert RP_{n}\right]\left[(CS_{q-1})^{2^{n}}\vert LP_{n}\right]\\
\left[(CS_{q-1})^{2^{n}}\vert RP_{n}\right]^{p-3}\end{array}\label{eq:th2Gpatt3}\end{equation}

\end{enumerate}
\item The pattern of the supercycle associated to the $p\cdot2^{n+1}$ last
appearance saddle-node orbit, within the $q$-window, by applying
Theorem~4 is computed as:

\begin{enumerate}
\item Repeat $p\cdot2^{n+1}$ times the pattern $CS_{q-1}$ \[
(CS_{q-1})^{p\cdot2^{n+1}}\]

\item Conjugate it according to (\ref{eq:th2Gpatt2})\begin{equation}
\begin{array}{r}
\left[(CS_{q-1})^{2^{n}}\vert CP_{n}\right]\left[(CS_{q-1})^{2^{n}}\vert RP_{n}\right]\left[(CS_{q-1})^{2^{n}}\vert LP_{n}\right]\\
\left[(CS_{q-1})^{2^{n}}\vert RP_{n}\right]^{2p-3}\end{array}\label{eq:th2Gpatt4}\end{equation}

\end{enumerate}
\end{itemize}
For $p\rightarrow\infty$, the patterns (\ref{eq:th2Gpatt3}) and
(\ref{eq:th2Gpatt4}) converge to the same Misiurewicz point with
preperiod\begin{equation}
\left[(CS_{q-1})^{2^{n}}\vert CP_{n}\right]\left[(CS_{q-1})^{2^{n}}\vert RP_{n}\right]\left[(CS_{q-1})^{2^{n}}\vert LP_{n}\right]\label{eq:misiurewicz-even-pre}\end{equation}
and period\begin{equation}
\left[(CS_{q-1})^{2^{n}}\vert RP_{n}\right]\label{eq:misiurewicz-even-per}\end{equation}

\item [(ii)]Let $n$ be odd.

The pattern of the supercycle associated to the first appearance $p\cdot2^{n}$
saddle-node orbit in the canonical window is (see \ref{eq:bandas-canonica_1})
\begin{equation}
CS_{p\cdot2^{n}-1}=CP_{n}LP_{n}RP_{n}(LP_{n})^{p-3}\label{eq:th2Gpatt5}\end{equation}
where $p=3,\,5,\,7,\,\dots$. Also, the pattern of the supercycle
associated to the last appearance $p\cdot2^{n+1}$ saddle-node orbit
in the canonical window is given by (see \ref{eq:bandas-canonica_2})\begin{equation}
CS_{p\cdot2^{n+1}-1}=CP_{n}LP_{n}RP_{n}(LP_{n})^{2p-3}\label{eq:th2Gpatt6}\end{equation}

\begin{itemize}
\item Pattern of the supercycle associated to the $p\cdot2^{n}$ first appearance
saddle-node orbit, within the $q$-window, by applying Theorem~4
is computed as:

After repeating $p\cdot2^{n}$ times the pattern $CS_{q-1}$ and conjugating
it according to (\ref{eq:th2Gpatt5}) we obtain\begin{equation}
\begin{array}{r}
\left[(CS_{q-1})^{2^{n}}\vert CP_{n}\right]\left[(CS_{q-1})^{2^{n}}\vert LP_{n}\right]\left[(CS_{q-1})^{2^{n}}\vert RP_{n}\right]\\
\left[(CS_{q-1})^{2^{n}}\vert LP_{n}\right]^{p-3}\end{array}\label{eq:th2Gpatt7}\end{equation}

\item Pattern of the supercycle associated to the $p\cdot2^{n+1}$ last
appearance saddle-node orbit, within the $q$-window, by applying
Theorem~4 is computed as:

After repeating $p\cdot2^{n+1}$ times the pattern $CS_{q-1}$ and
conjugating it according to (\ref{eq:th2Gpatt6}) we obtain\begin{equation}
\begin{array}{r}
\left[(CS_{q-1})^{2^{n}}\vert CP_{n}\right]\left[(CS_{q-1})^{2^{n}}\vert LP_{n}\right]\left[(CS_{q-1})^{2^{n}}\vert RP_{n}\right]\\
\left[(CS_{q-1})^{2^{n}}\vert LP_{n}\right]^{2p-3}\end{array}\label{eq:th2Gpatt8}\end{equation}

\end{itemize}
For $p\rightarrow\infty$, the patterns (\ref{eq:th2Gpatt7}) and
(\ref{eq:th2Gpatt8}) converge to the same Misiurewicz point with
preperiod\begin{equation}
\left[(CS_{q-1})^{2^{n}}\vert CP_{n}\right]\left[(CS_{q-1})^{2^{n}}\vert LP_{n}\right]\left[(CS_{q-1})^{2^{n}}\vert RP_{n}\right]\label{eq:misiurewicz-odd-pre}\end{equation}
and period\begin{equation}
\left[(CS_{q-1})^{2^{n}}\vert LP_{n}\right]\label{eq:misiurewicz-odd-per}\end{equation}
$\blacktriangleleft$

\end{enumerate}
\textbf{Let the $R$-parity of the $q$-periodic supercycle pattern
be odd}. Given that the $R$-parity of $CS_{q-1}$ is odd, the conjugated
patterns $\overline{CS_{p\cdot2^{n}-1}}$ and $\overline{CS_{p\cdot2^{n+1}-1}}$
will be used on applying Theorem~4. Hence, the results obtained will
be the same.

Corollary to Theorem~3 is now rewritten, valid for any periodic window
in the bifurcation diagram.

\begin{description}
\item [{Corollary~to~Theorem~6.}] \emph{Band merging points in a $q$-periodic
window, whose patterns are given in (\ref{eq:misiurewicz-even-pre})-(\ref{eq:misiurewicz-even-per})
and (\ref{eq:misiurewicz-odd-pre})-(\ref{eq:misiurewicz-odd-per})
as the limit orbits of saddle-node orbit sequences, form a Misiurewicz
point cascade. Furthermore, this Misiurewicz point cascade converges
to the Myrberg-Feigenbaum point of that window, driven by the behavior
of the saddle-node bifurcation cascades, described in \cite{SanMartin07}.}
\end{description}
It is important to notice that the symbolic sequences of the Misiurewicz,
given by (\ref{eq:misiurewicz-even-pre}-\ref{eq:misiurewicz-even-per})
and (\ref{eq:misiurewicz-odd-pre}-\ref{eq:misiurewicz-odd-per}),
show the band merging points in any window, that is, not only those
(already known) corresponding to the canonic window, but also those
located inside every subwindow.

\section{\label{sec:disc-and-conclusions}Discussion and conclusions}

It has been shown, in the preceding sections, how the patterns of
those periodic orbits of the Feigenbaum cascade determine the patterns
of the orbits of the saddle-node bifurcation cascades. The accumulation
points corresponding to these latter patterns are the band merging
points, that are also Misiurewicz points. The sequence of these Misiurewicz
points converges to the Myrberg-Feigenbaum point, and determines the
chaotic band structure of the bifurcation diagram. This chain of relations
shows how the periodic part of a window determines its chaotic part:
the hyperbolic components of the set, determine the non-hyperbolic
ones. 

The relationship between the Feigenbaum cascade and the saddle-node
bifurcation cascade could be expected, because both cascades share
the same scaling law found by Feigenbaum \cite{Feigenbaum78}. On
the other hand, this scaling is the direct consequence of the renormalization
equation. This equation has only one fixed point in the functional
space, whose stable manifold is one-dimensional. Thus, Feigenbaum's
cascade and saddle-node cascade can but pertain to the same scaling
universality \cite{deMelo93}.

The problem of localizing the chaotic band merging points, as well
as the convergence of that sequence of points, is a problem largely
addressed numerically by many authors as, for instance, by Hao and
Zhang \cite{Hao82} in their {}``Bruselator oscillator'' model.
Both problems can be approached using the Corollary to Theorem~6,
stated above, and the saddle-node cascade convergency, described in
\cite{SanMartin07}. 

From the symbolic sequences of a pitchfork orbit in the Feigenbaum's
cascade, the symbolic sequences of the supercycles of the saddle-node
orbits can be computed. After which, Myrberg's formula \cite{Myrberg63}
can be used to compute the value of the parameter giving rise to those
orbits. This circumstance can be very useful to the experimenter,
because once the Feigenbaum cascade has been observed, the corresponding
saddle-node bifurcations can be spotted from it. Given that many dynamical
systems, in particular lasers, present period doubling cascades, it
easy for the experimenter to explicitly localize the saddle-node bifurcation
cascades from the experimentally obtained return-map \cite{Pisarchik00}
. As a result, it is also possible to spot the intermittency cascades
associated with that saddle-node bifurcation cascade. We must take
into account that without an exact knowledge of the parameter values,
it is practically impossible to localize these cascades, because of
the geometrical stretching of those regions under saddle-node bifurcations.

We want to point out other possible uses from an experimental point
of view. A relevant fact for the experimenter is the possibility to
assess the localization of the Misiurewicz points which separate chaotic
bands within any window. In this way, parameter values can be bound
to the particular region of interest where the sought phenomena take
place. 

There is yet another interesting point. In all the results just obtained,
orbit sequences have been considered. Each of them had a symbolic
sequence associated to it. These symbolic sequences on their own,
can be used to assess the underlying universal scaling laws \cite{LopezRuiz06}.
This can be reflected, for instance, in the fact that the previously
signaled Misiurewicz points scale according to the same Feigenbaum
scaling scheme.

All that has been shown in this paper is immediately extendible to
unimodal maps with extrema of the form $x^{2n}$ under very general
considerations.

\appendix

\section*{Appendix}

\begin{description}
\item [{Lema~1}] \emph{Let $f(x;\,\mu)$ be an unimodal $C^{2}$ class
map of the interval $[a,\, b]$ into itself, such that its critical
point $x_{c}\in(a,\, b)$ is a maximum (minimum). If for $\mu=\mu_{0}$,
there exists a $q$-periodic supercycle having a pattern $CS_{q-1}$,
then $f^{q}(x_{c};\,\mu_{0})$ has a maximum (minimum) when the $R$-parity
of $CS_{q-1}$ is even, while it has a minimum (maximum) in case of
odd $R$-parity of $CS_{q-1}$.}
\item [{Proof:}] Let us consider a value of $\mu$ in a small neighborhood
of $\mu_{0}$, such that $f^{q}(x;\,\mu)$ has a fixed point at some
$x=x_{c}+\varepsilon$, for some arbitrarily small $\varepsilon>0$.
At $x=x_{c}$, $f^{q}(x_{c};\,\mu)$ has a critical point, given that
$f(x_{c};\,\mu)$ has a critical point, for $x=x_{c}$. To determine
whether it is a maximum or a minimum, it is enough to see whether,
in a neighborhood of $x_{c}+\varepsilon$, the iterated function $f^{q}(x_{c}+\varepsilon;\,\mu)$
is increasing (corresponding to $f$ having a minimum at $x_{c}$)
or decreasing (corresponding to a maximum). That is, whether $\left[f^{q}\right]^{\prime}(x_{c}+\varepsilon;\,\mu)$
is positive or negative.

\[
\left[f^{q}\right]^{\prime}(x_{c}+\varepsilon;\,\mu)=\prod_{i=1}^{q}f^{\prime}(x_{i})\]
where the $x_{i}$ represent each of the $q$ iterates of the function
$f$ period, starting at $x_{1}=x_{c}+\varepsilon$. It turns out
that\[
\mbox{sign}\left[f^{q}\right]^{\prime}(x_{c}+\varepsilon;\,\mu)=\prod_{i=1}^{q}\mbox{sign }f^{\prime}(x_{i};\,\mu)\]
Let us assume $f$ to have a maximum at $x_{c}$, that is, to be an
increasing function in $(a,\, x_{c})$ (i.e. with positive derivative)
and a decreasing function in $(x_{c},\, b)$ (i.e. with negative derivative).
When $f^{q}$ has a minimum at $x_{c}$ (i.e. $\mbox{sign}\left[f^{q}\right]^{\prime}(x_{c}+\varepsilon;\,\mu)$
is positive) the number of $x_{i}$ laying in $(x_{c},\, b)$ will
be even, therefore, the number of $R$-symbols in the pattern of the
orbit will be even. Conversely, when $f^{q}$ has a maximum at $x_{c}$,
the number of $R$-symbols in the pattern will be odd.

Considering $x_{1}=x_{c}+\varepsilon\rightarrow x_{c}$, we get back
the superstable orbit, $CS_{q-1}$. This represents the deletion of
the first $R$ of the pattern, corresponding to $x_{1}$. Based on
the continuous variation of the function with $\mu$, we find that
for an even $R$-parity of the supercycle pattern, a maximum will
be present at $x=x_{c}$, while for an odd $R$-parity, there will
be a minimum. $\blacktriangleleft$

\item [{Note~1:}] We need to know the $R$-parity of $CP_{n}$ in order
to compute F-harmonics of the pitchfork orbits in the proof. It is
clear that for the period-$1$ orbit supercycle, whose sequence is
$C$, its first pitchfork has the sequence $CR$ and odd $R$-parity;
the orbit originated by the second pitchfork has sequence $CR[C]R\rightarrow CRLR$,
as given by the Metropolis-Stein-Stein harmonic \cite{MetropolisSteinStein73},
having even $R$-parity; the iteration of these harmonics gives odd
$R$-parities for those $CP_{n}$ with $n$ odd, while those $CP_{n}$
with even $n$, have even $R$-parity. As a consequence of this, it
is easy to see that $\overline{CP_{n}}$ has odd (even) $R$-parity
whenever $n$ is even (odd).
\item [{Lemma~2}] \emph{The $n$-th pitchfork bifurcation of $CS_{q-1}$,
that is, $CP_{n,q}$ is given by\[
\left((CS_{q-1})^{2^{n}}\vert CP_{n}\right)\mbox{ for even }R\mbox{-parity of }CS_{q-1}\]
\[
\left((CS_{q-1})^{2^{n}}\vert\overline{CP_{n}}\right)\mbox{ for odd }R\mbox{-parity of }CS_{q-1}\]
Furthermore, $CP_{n,q}$ has the same $R$-parity as $CP_{n}$ (respectively,
$\overline{CP_{n}}$) for even (respectively, odd) $R$-parity of
$CS_{q-1}$.}
\item [{Proof.}]~

\textbf{Let the $R$-parity of $CS_{q-1}$ be even.} When the MSS
composition rule \cite{MetropolisSteinStein73} is applied, the sequence
$S_{q-1}$ does not play any role, because it does not change the
$R$-parity of the new pattern. Only the $C$'s formerly changed have
a role. Therefore we work as if with the $n$-th pitchfork bifurcation
of the period-$1$ orbit.

\textbf{Let the $R$-parity of $CS_{q-1}$ be odd.} The proof is similar
to the former point, but the sequence $S_{q-1}$ changes the $R$-parity
of the new pattern when the composition rule is applied. Therefore
we work as if with the conjugated of the $n$-th pitchfork bifurcation
of the period-$1$ orbit.

Whatever the $R$-parity of $S_{q-1}$ is, the $(CS_{q-1})^{2^{n}}$
sequence always has an even number of $S_{q-1}$ sequences, therefore
this even number of $S_{q-1}$ gives an even $R$-parity. So, the
$R$-parity of $\left((CS_{q-1})^{2^{n}}\vert CP_{n}\right)$ comes
from the $C$'s inserted from $CP_{n}$, that is, the $R$-parity
of $\left((CS_{q-1})^{2^{n}}\vert CP_{n}\right)$ is that of $CP_{n}$.
In the same way it is proven that $\left((CS_{q-1})^{2^{n}}\vert\overline{CP_{n}}\right)$
has the same $R$-parity as $\overline{CP_{n}}$.

\end{description}


\begin{thebibliography}{10}

\bibitem{Feigenbaum78}
M.~J. Feigenbaum.
\newblock Quantitative universal for a class of nonlinear transformations.
\newblock {\em J. Statist. Phys.}, 19:25--, 1978.

\bibitem{Feigenbaum79}
M.~J. Feigenbaum.
\newblock The universal metric properties of nonlinear transformations.
\newblock {\em J. Statist. Phys.}, 21:669--706, 1979.

\bibitem{Milnor88}
J.~Milnor and W.~Thurston.
\newblock On iterated maps of the interval.
\newblock In {\em Lect. Notes Math.}, number 1342, pages 465--563.
  Springer-Verlag, Berlin, New York, 1988.

\bibitem{SanMartin07}
J.~San~Mart{\'\i}n.
\newblock Intermittency cascade.
\newblock {\em Chaos, Solitons and Fractals}, 32:816--831, 2007.

\bibitem{Misiurewicz91}
M.~Misiurewicz and Z.~Nitecki.
\newblock Combinatorial patterns for maps of the interval.
\newblock {\em Mem. Am. Math. Soc.}, 94:456, 1991.

\bibitem{Churukian96}
A.~D. Churukian and D.~R. Snider.
\newblock Finding the windows of regular motion within the chaos of ordinary
  differential equations.
\newblock {\em Phys. Rev. E}, 53:74--79, 1996.

\bibitem{Xu96}
D.~Xu and S.~R. Bishop.
\newblock Switching between orbits in a periodic window.
\newblock {\em Phys. Rev. E}, 54:6940--6943, 1996.

\bibitem{Bishop98}
S.~R. Bishop, D.~Xu, C-Y Liau, and E-S Chan.
\newblock Applying control in periodic windows.
\newblock {\em Chaos, Solitons and Fractals}, 9:1297--1305, 1998.

\bibitem{Blackburn98}
J.~A. Blackburn and N.~Gronbech-Jensen.
\newblock Phase diffusion in a chaotic pendulum.
\newblock {\em Phys. Rev. E}, 53:3068--3072, 1996.

\bibitem{Derrida79}
B.~Derrida, A.~Gervois, and Y.~Pomeau.
\newblock Universal metric properties of bifurcations of endomorphisms.
\newblock {\em J. Phys A}, 12:269--296, 1979.

\bibitem{MetropolisSteinStein73}
N.~Metropolis, M.~L. Stein, and P.~R. Stein.
\newblock On finite limit sets for transformations on the unit interval.
\newblock {\em J. Comb. Theory A}, 15:25--44, 1973.

\bibitem{Pastor96}
G.~Pastor, M.~Romera, and F.~Montoya.
\newblock On the calculation of {M}isiurewicz patterns in one-dimensional
  quadratic maps.
\newblock {\em Physica A}, 232:536--553, 1996.

\bibitem{Romera96}
M.~Romera, G.~Pastor, and F.~Montoya.
\newblock On the cusp and the tip of a midget in the {M}andelbrot set antenna.
\newblock {\em Phys Lett A}, 221:158--162, 1996.

\bibitem{Pastor02}
G~Pastor, M.~Romera, G.~Alvarez, and Montoya F.
\newblock Operating with external arguments in the {M}andelbrot set antenna.
\newblock {\em Physica D}, 171:52--71, 2002.

\bibitem{Pastor03}
G.~Pastor, M.~Romera, G.~Alvarez, and F.~Montoya.
\newblock How to work with one-dimensional quadratic maps.
\newblock {\em Chaos, Solitons and Fractals}, 18:899--915, 2003.

\bibitem{Post91}
T.~Post and H.W. Capel.
\newblock Windows in one-dimensional maps.
\newblock {\em Phys. A}, 178:62--100, 1991.

\bibitem{Myrberg63}
P.~J. Myrberg.
\newblock Iteration der reellen polynome zweiten grades {III}.
\newblock {\em Ann. Acad. Sci. Fenn., Series A}, 336:1--18, 1963.

\bibitem{Zheng84}
W.-Z. Zheng, B.-L. Hao, G.-R. Wang, and S.-G. Chen.
\newblock Scaling property of period n-tupling sequences in one-dimensional
  mappings.
\newblock {\em Commun. Theor. Phys.}, 3:283--295, 1984.

\bibitem{Hunt99}
B.~R. Hunt, J.~A.~C. Gallas, C.~Grebogi, J.~A. Yorke, and H.~Kocak.
\newblock Bifurcation rigidity.
\newblock {\em Physica D}, 129:35--56, 1999.

\bibitem{deMelo93}
W.~de~Melo and S.~van Strein.
\newblock {\em One-dimensional dynamics}.
\newblock Springer-Verlag, Berlin, 1993.

\bibitem{Hao82}
B.-L. Hao and S.-Y. Zhang.
\newblock Hierarchy of chaotic bands.
\newblock {\em J. Stat. Phys.}, 28:769--, 1982.

\bibitem{Pisarchik00}
A.~N. Pisarchik, R.~Meucci, and F.~T. Arecchi.
\newblock Discrete homoclinic orbits in a laser with feedback.
\newblock {\em Phys. Rev. E}, 62:8823--8825, 2000.

\bibitem{LopezRuiz06}
R.~López-Ruiz.
\newblock Order in binary sequences and the routes to chaos.
\newblock {\em Chaos, Solitons and Fractals}, 27:1316--1320, 2006.

\end{thebibliography}
\end{document}